\documentclass[preprint,
showpacs,
preprintnumbers,
amsmath,
amssymb,
prl,
superscriptaddress]{revtex4-1}

\usepackage[usenames,dvipsnames]{color}
\usepackage{graphicx}
\usepackage{dcolumn}
\usepackage{simplewick}
\usepackage{bm}
\usepackage{hyperref}
\usepackage{amsmath}

\usepackage{natbib}

\newcommand{\PTO}{{\rm PbTiO$_3$}}
\newcommand{\BTO}{{\rm BaTiO$_3$}}

\begin{document}

\title{Enhancement of electrocaloric response through quantum effects}

\author{P. Jouzdani}
\author{S. Cuozzo}
\author{S. Lisenkov}
\author{I. Ponomareva}

\affiliation{Department of Physics, University of South Florida, Tampa, 
             Florida 33620, USA}

\begin{abstract}
A semiclassical approach that incorporates quantum mechanical behavior of heat capacity in direct caloric effect simulations is proposed. Application of this methodology to study electrocaloric effect in prototypical ferroelectrics \PTO\, and \BTO\, reveals severe underestimation of electrocaloric response at lowest temperatures by classical simulations. The discrepancy between semiclassical and classical results are found to be largest  in ferroics with Debye temperature exceeding the Curie point. A route to enhance electrocaloric effect by tuning the Debye temperature in composite materials is proposed.

\end{abstract}


\maketitle

The electrocaloric effect has received a lot of attention recently \cite{Rev-Scott,Rev-Mayo,Rev-alpay_mantese_trolier-mckinstry_zhang_whatmore_2014} owing to its potential for solid state refrigeration technology \cite{Rev-ValantECEFutureSolid,Rev-BlumenthalClassificationECECooling} and  success in applying large electric fields to thin ferroic samples \cite{Rev-MischenkoGiantECEThinFilm}. It is defined as a reversible change in temperature under adiabatic application or removal of  the electric field. Alternatively, it can be defined as a reversible isothermal entropy change under the application or removal of the external field. Its potential for solid state cooling - environmentally friendly and energy efficient alternative to conventional refrigeration - largely relies on the success in discovery of materials or structures with technologically significant electrocaloric change in temperature achieved by an application of relatively low electric fields or voltages. In addition, very recently, the electrocaloric effect was proposed to enhance or establish magnetoelectric coupling in magnetoelectrics \cite{Rev-ChangThermallyMaechMagnetoelectricPRL}.  
The electrocaloric change in temperature can be either directly measured under  adiabatic application of the field or estimated with the help of the Maxwell relations \cite{Rev-YangLiuDirectIndirect}. These two approaches are called the direct and indirect methods, respectively. Very recently, the success in the development of atomistic simulations that mimic the direct method \cite{Rev-LisenkovPRBIntrinsicECE,Rev-PonomarevaPRLBridging,Rev-TakeshiDirectMolecularDynamics,Rev-LisenkovECECarbonNano} has led to many insights into the caloric effects. Some examples include predictions of giant elastocaloric effect in ferroelectric ferroelastics \cite{Rev-LisenkovElastocaloricPRB} which  was later confirmed experimentally \cite{Rev-ChauhanECECeramics} and prediction of multicaloric effect in ferroelectric ferroelastics \cite{Rev-LisenkovMulticaloricPRB} which has also received experimental confirmation recently \cite{Rev-ChauhanMulticaloric}. 
The power of electrocaloric effect (ECE) simulations, however, is not limited to the predictions of novel effects and atomistic insights. They often allow to extend the available experimental data to temperatures or fields  for which the data are challenging to obtain. At the same, to the best of our knowledge, all current computational techniques utilize the classical framework to compute the caloric effects. While the classical approach is often well justified by the classical nature of the effect, there exists one potential source of quantum mechanical corrections which has so far been overlooked. Caloric effect depends strongly on the heat capacity of the material, which in turn, shows quantum mechanical features, typically below the Debye temperature.  This raises several questions. What  role do these quantum contributions play in the ECE and is it possible to quantify these contributions? How reliable are classical simulations below the Debye temperature and under what conditions they provide an accurate description of the effect? Is it possible to develop a computational tool that could address these questions? Finally, could the quantum contributions be exploited to enhance or engineer  the caloric effect?   

The goal of this Letter is to address  these questions. In particular, in this Letter we: i) propose a  computational framework that allows accurate simulations of heat capacity in adiabatic computations including the quantum regime; ii) apply the methodology to the case of prototypical ferroelectrics to establish the role of quantum effects in electrocaloric response  and the applicability of classical simulations; iii) predict a route to enhance the caloric effects by taking advantage of the quantum regime.       

To achieve our methodological goal of incorporating quantum mechanical heat capacity in the adiabatic simulations we turn to our original approach for the direct caloric effect simulations in ferroics \cite{Goal-LisenkovPRBIntrinsicECE,Goal-PonomarevaPRLBridging}. In such an approach the direct simulations of the caloric effects are achieved through the adiabatic Monte Carlo technique which introduces additional degrees of freedom, called demons, that carry energy associated with conjugate momentum in microcanonical formulation. The demons are allowed to exchange energy with the lattice, however, are not allowed to create or annihilate energy, thus maintaining the adiabatic conditions. In other words, their function is to redistribute the energy between different components of a closed system. Application of an external field (electric, magnetic or stress field) changes the potential energy of the lattice and initiates energy flow between the potential and kinetic energies. The latter one is modeled by the demons each of which carry $k_BT$ of energy. The change in the  average energy of the demons is then used to compute the change in temperature due to the application of the external field. In classical simulations the number of demons could be related to the number of degrees of freedom in the system, $N$, as $n_{dem}=N/2$ under the assumption that each demon carries  $k_BT$ energy.  For example, for the case of perovskites, which are  the focus of this work, $N=15$ per 5 atom unit cell and  the associated heat capacity per unit cell 
$C \approx n_{dem}k_B+\frac{N}{2}k_B=Nk_B=21\times10^{-23}$~J/K (or 124.7 J/mol$\cdot$K)  is in excellent agreement with the high temperature value of 118.9 J/mol$\cdot$K  in \BTO\,  \cite{Kallaev2013} and of 124.7 J/mol$\cdot$K in \PTO\, \cite{ PTOExperimentalRossetti}.

\begin{figure}[h]
  \includegraphics[width=1.0\textwidth, angle=-0]{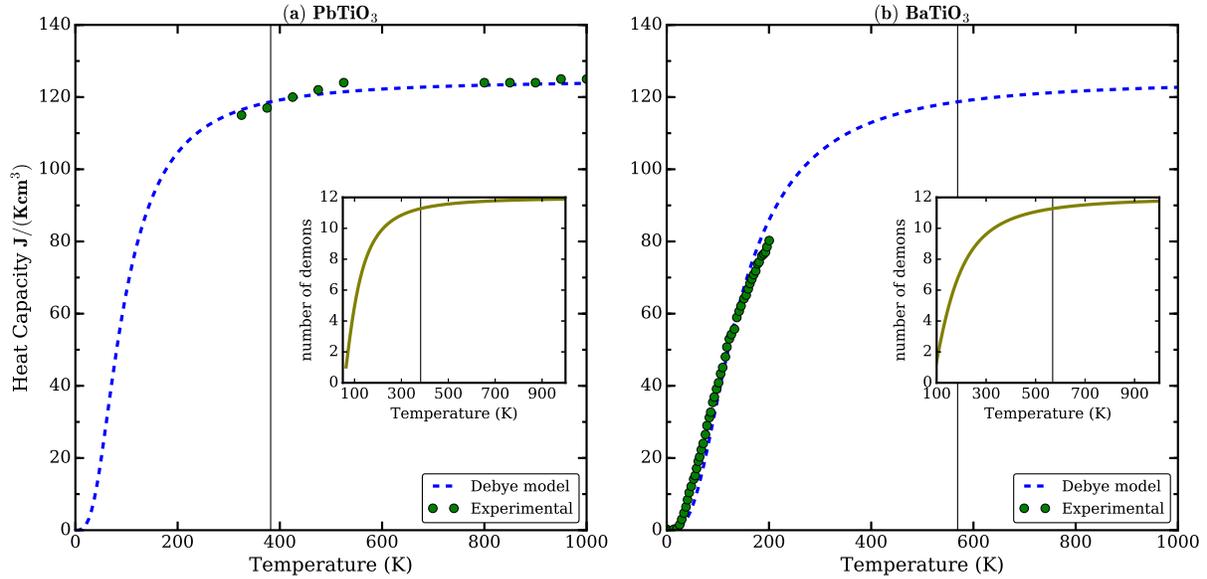}
  \caption{(color online) Heat capacity in \PTO\, (a) and \BTO\, (b). Symbols give experimental data from the literature, while dashed lines give the fit with the Debye model. Solid vertical line gives the Debye temperature obtained from the fit. Note, that we removed points in the vicinity of the transition temperatures to simplify the fitting. Experimental data for  \PTO\, are taken from Ref. \onlinecite{PTOExperimentalRossetti}, while for  \BTO\, the data are taken from Ref. \onlinecite{doi:10.1021/ja01128a054}.  Insets give the number of demons used in the simulations as a function of temperature. }
  \label{Fig1}
\end{figure}

To incorporate quantum mechanical heat capacity into adiabatic Monte Carlo technique we turn to the experimental data for the heat capacity in perovskites. Fig. \ref{Fig1} gives the experimental data for the heat capacity of \PTO\, \cite{PTOExperimentalRossetti} and \BTO\, \cite{doi:10.1021/ja01128a054}  along with the fit to the data with the Debye heat capacity model $C_{Debye}=C_{D-P}(\frac{\Theta_D}{T})^3\int ^\frac{\Theta_D}{T} _0  \frac{x^4e^x}{(e^x-1)^2}dx$, where the Dulong-Petit heat capacity limit $C_{D-P}$  and the Debye temperature $\Theta_D$ are two adjustable parameters. The model yields $C_{D-P}=124.7$ J/mol$\cdot$K for both materials and the Debye temperatures of $349$~K and $569$~K for \PTO\, and \BTO,  respectively. The Debye temperatures for \PTO\, and \BTO\, from the literature are $335$~K and $513$~K, respectively \cite{DebyeTempPTO, DebyeTempBTO}. Having established that the heat capacity in perovskite \PTO\, and \BTO\, follow the Debye model we incorporate this model in our adiabatic Monte Carlo simulations. The approach is to allow the number of demons to vary with temperature to satisfy the Debye heat capacity per unit cell, $C_{Debye}=\langle n_{dem} \rangle k_B+\frac{\partial U_{pot}}{\partial T}$, where $\langle n_{dem} \rangle $ is the thermal average for the number of demons at a given temperature and the last term describes the classical potential energy contribution to the heat capacity. This semiclassical approach yields the following expression for the number of demons per unit cell

\begin{equation}
  \label{eq:ndem}
  \langle n_{dem} \rangle k_B = C_{D-P}\left (\frac{\Theta_D}{T} \right )^3\int ^{\Theta_D/T}_0  \frac{x^4e^x}{(e^x-1)^2}dx - \frac{\partial U_{pot}}{\partial T}
\end{equation}

 Implementation of the temperature dependent number of demons  from Eq.(\ref{eq:ndem}) in simulations yields the heat capacity given by the dashed line in Fig.\ref{Fig1} and that satisfies both the Debye model and the experimental data. Near the phase transition temperature the heat capacity diverges owing to the potential energy contribution in  Eq.(\ref{eq:ndem}), however, this is not shown in Fig.\ref{Fig1}. Note, that the thermal average number of demons is not necessarily an integer. Technically, a noninteger $<n_{dem}>$ is modeled through carrying out two simulations for the two integer numbers of demons that bracket $<n_{dem}>$ and then computing weighted average for the property obtained from both simulations
\footnote {For example, for the case of $<n_{dem}> = x a + (1-x) b$, where $a$ and $b$ are the two integers that bracket $<n_{dem}>$ the weighted average property $<A>$ is computed as $<A>= x A_a+(1-x)A_b$, where $A_a$ and $A_b$ are the property $A$ obtained from simulations with $a$ and $b$ numbers of demons, respectively. }. At  low temperatures (typically below 100 K) Eq.(\ref{eq:ndem})  yields negative $<n_{dem}>$ due to the classical contribution of the potential energy to the heat capacity. For these temperatures the lowest number of demons of one could be used. However, in this study we choose not to report the data for such temperatures. It should be noted that, while we chose the Debye model to reproduce experimental data for the heat capacity,  the proposed methodology is general and can be used with any other model for the heat capacity or even empirical values, which will replace the Debye integral in Eq.(\ref{eq:ndem}). The advantage of our approach is that the desired value of the heat capacity can be accurately reproduced in adiabatic simulations by adjusting the number of demons.

We begin by an application of this approach to study the ECE in prototypical ferroelectrics  \PTO\, and \BTO. Both materials are simulated using  a supercell of 12x12x12 unit cells periodic along the three Cartesian directions to model the bulk sample \footnote{Some data were cross checked by using a simulation supercell of 16x16x16 unit cells.}. The potential energy of the supercell is given by the first-principles-based effective Hamiltonians of Refs.\cite{ManiAtomisticPRB,herchignw}. The degrees of freedom for the effective Hamiltonian \cite{ZhongPRBHamiltonian} include local soft modes which are proportional to the local dipole moment in the unit cell and strain variables which describe unit cell and supercell deformations. The effective Hamiltonian includes energies that describe on-site, short-range and long-range interactions between the local modes, elastic deformations and the coupling between the degrees of freedom that describes electrostriction and piezoelectricity. The effective Hamiltonian reproduces well the sequence of the phase transitions and the associated transition temperatures  in both \PTO\, and \BTO. The supercells are first annealed from temperatures well above the Curie point down to the temperatures of 50 K in steps of 5 K using the Metropolis Monte Carlo technique. For each temperature we use 3$\cdot$10$^{5}$ Monte Carlo sweeps for \BTO\  and 8$\cdot$10$^{4}$ Monte Carlo sweeps for \PTO\,  to obtain equilibrated structures. At the next step the equilibrated supercells are subjected to an external electric field under adiabatic conditions. In ferroelectric phases the electric field is applied along the direction of the polarization, while in paraelectric phases the field is applied along [001] direction. Such field can not induce a phase transition, which is known to have a dramatic effect on the ECE \cite{Rev-PonomarevaPRLBridging,Rev-TakeshiDirectMolecularDynamics,PhysRevB.96.014102} Adiabatic conditions are simulated using our semiclassical adiabatic Monte Carlo with the heat capacity described by the Debye model. At each temperature an applied electric field is slowly increased from 0 to 1000~kV/cm at the rate of $0.01$~kV/cm per one Monte Carlo sweep and then reduced to zero at the same rate to check for reversibility. At each temperature the number of demons is computed from Eq.(\ref{eq:ndem}) but kept constant throughout the entire simulation since the change in temperature is usually not large enough to significantly change this number. The temperature during the application and removal  of the electric field is computed from the average energy of all demons.  The actual number of demons used in the simulations is given in the insets to Fig.\ref{Fig1}. In the high temperature classical limit $n_{dem}=12$ which corresponds to the difference between the high-temperature Dulong-Petit heat capacity value of 15k$_B$ and the potential energy contribution of $\frac{\partial U_{pot}}{\partial T}\approx$3k$_B$  computed from the effective Hamiltonian. The latter one is smaller than the classical potential energy contribution of (15/2)k$_B$  per unit cell  due to the reduced number of degrees of freedom used in the Hamiltonian \cite{ZhongPRBHamiltonian}.  
To investigate  the role that quantum corrections play in the ECE we also carried out classical ECE simulations using the number of demons that correspond to the high temperature Dulong-Petit value of heat capacity.

\begin{figure}[h]
  \includegraphics[width=1.0\textwidth, angle=-0]{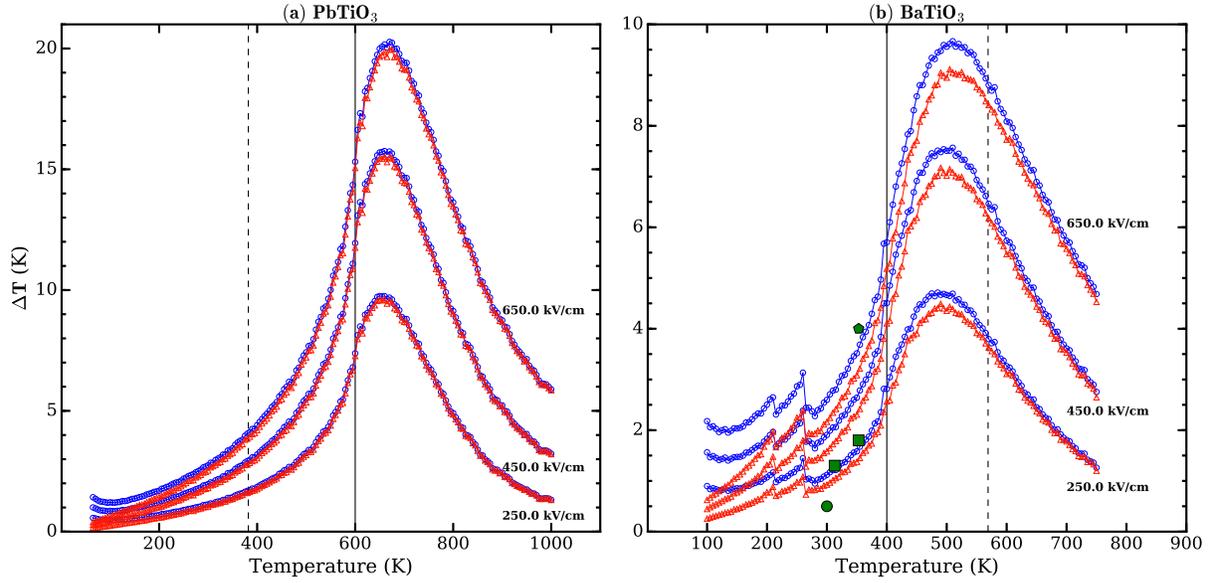}
 
  \caption{ (color online) Temperature evolution of the electrocaloric change in temperature in \PTO\, (a) and \BTO\, (b) under different applied electric fields computed using classical (triangles) and semiclassical (empty circles)  approaches. Solid vertical lines indicate computational  Curie temperature, while dashed vertical lines give computational Debye temperature. Squares  give experimental data from the literature for the electric field of 176~kV/cm, while pentagon gives data for the electric field of 352~kV/cm \cite{ExperimentalBTO}. Filled circle gives experimental data for the electric field of 300~kV/cm \cite{0022-3727-43-3-032002}.}
   \label{Fig2}
\end{figure}

Fig.\ref{Fig2} shows the temperature evolution of the  electrocaloric change in temperature, $\Delta T (T)$, under different electric fields obtained from both classical and semiclassical computations. We have also added some experimental data from the literature for \BTO\, to Fig.\ref{Fig2} which indicate good agreement between our predictions and experimental measurements. While we could not find experimental data for \PTO\, our room temperature computational value of 0.3~K under the electric field of 67~kV/cm compares well with experimentally measured response of 0.1~K in Pb(Zr$_{0.2}$Ti$_{0.8}$)O$_3$ samples \cite{PhysRevB.90.094116} under the same conditions.  We first notice that in all cases classical simulations underestimate the ECE below the Debye temperature. However, the most drastic differences occur at the lowest investigated temperatures. For example, in case of \BTO\, at 100 K we find more than three times the increase in $\Delta T$ with respect to the classical estimates. Moreover, even the qualitative behavior of $\Delta T (T)$ changes at the lowest temperatures. While the classical $\Delta T$ decreases monotonically with $T$, the semiclassical $\Delta T$ passes through a minimum and increases at the lowest reported temperatures. This trend is suggestive of a divergent behavior on approaching zero Kelvin which we tentatively attribute to the rapid decrease in heat capacity which overpowers the decrease in the pyroelectric coefficient. Overall, we find that classical simulations underestimate electrocaloric $\Delta T$ from  6\% to 71\%  on decreasing the temperature from $\Theta_D$ to $\Theta_D/5$, independent of the  strength of the electric field. This means that the semiclassical $\Delta T$ is twice the classical estimate at temperatures  around $\Theta_D/3$. At  $3\Theta_D/2$ the underestimation is only 3\% as the heat capacity approaches its high temperature limit. This explains why classical and semiclassical results merge for \PTO\, above 500~K and for \BTO\, above 770~K (see Fig.\ref{Fig2}).  Note, that the discontinuities in $\Delta T$ of \BTO\, at 210 and 260 K are due to the first order phase transitions associated with the discontinuity in entropy.  They have been previously reported from experiments \cite{PhysRevB.82.134119} and computations \cite{Rev-PonomarevaPRLBridging}. It should be noted that in our simulations we assumed constant Debye temperature. However, in some ferroelectrics the Debye temperature may differ between different phases. For example, in \BTO\, the Debye temperature in tetragonal phase was reported to be  14\% smaller than in the paraelectric phase \cite{Ledbetter_Austin_Kim_Lei_1987}. To verify the effect that the difference in Debye temperature might have on our results we repeated the ECE calculations for \BTO\, using the tetragonal phase Debye temperature of 429~K \cite{Ledbetter_Austin_Kim_Lei_1987} which is 24\% lower than the one used in the original simulations. We found that in the tetragonal phase at room temperature the electrocaloric $\Delta T$ decreased on average by 10\% only. This finding suggests that our results are robust against variations of Debye temperature between different phases and, more importantly, the accuracy of the predictions can be systematically improved, thanks to the generality of our computational methodology.

Fig.\ref{Fig2} suggests one interesting observation. The difference between the classical and semiclassical results is much larger in case of \BTO\, as compared to the case of \PTO. To understand the origin of this we recall the Maxwell relation for the electrocaloric coefficient  $dT/dE=-\frac{T}{C_E}\left ( \frac{\partial P}{\partial T} \right )_E$, where $C_E$ is the heat capacity under the constant electric field $E$, and $P$ is the polarization. The expression predicts that the largest $dT/dE$ occurs when the pyroelectric coefficient $\left ( \frac{\partial P}{\partial T} \right )_E$ reaches extremal value, which at low electric fields, is in the vicinity of the Curie point. On the other hand, $dT/dE$ can be further enhanced through a decrease in the heat capacity which occurs below the Debye temperature. These arguments suggest that the largest deviation from classical predictions occur when the Debye temperature is higher than the Curie temperature of the material. This is also the case when we would expect the strongest enhancement of the ECE due to quantum contributions.

\begin{figure}[h]
  \includegraphics[width=\textwidth, angle=-0]{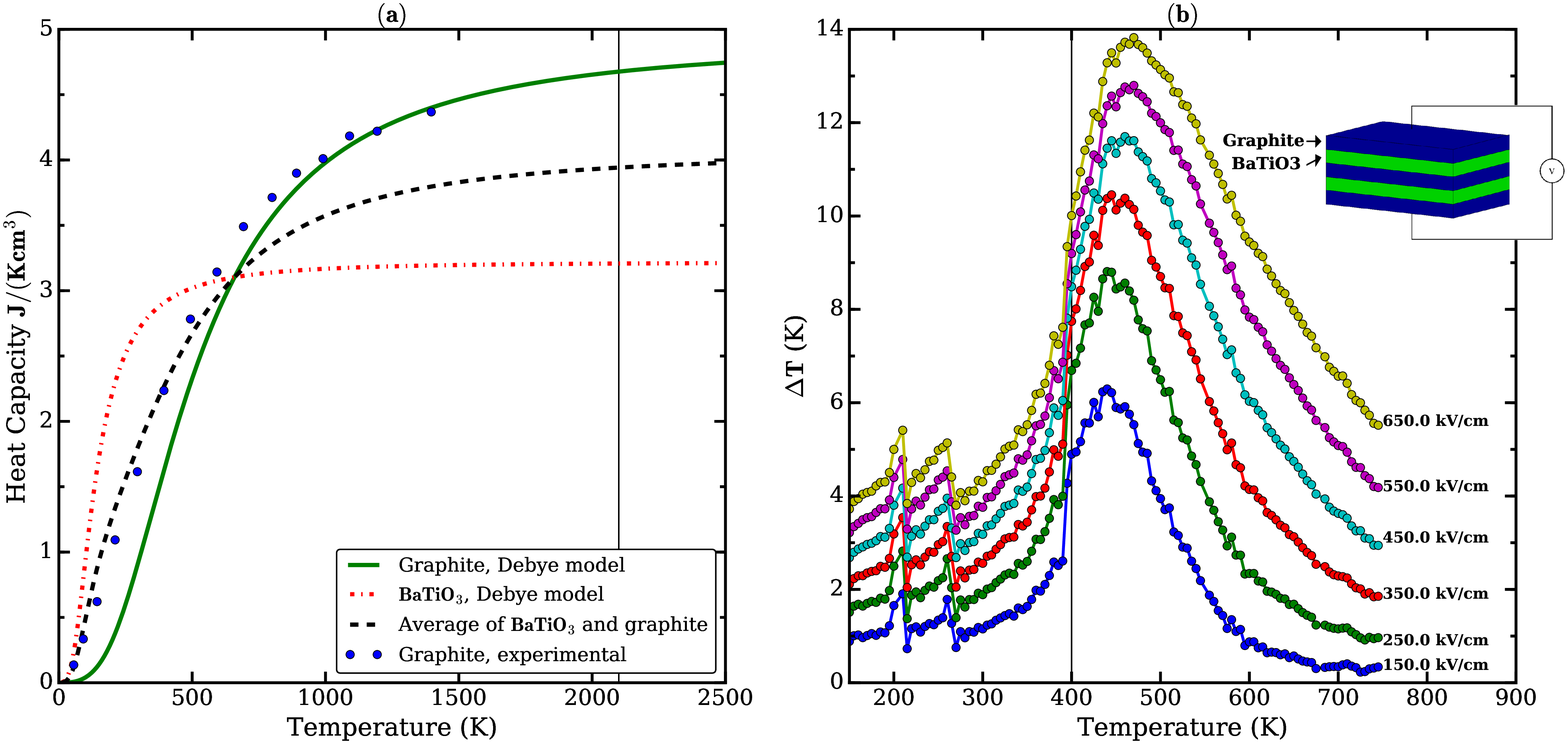}
  \caption{ (color online) Heat capacity of \BTO, graphite and their average one (a). Experimental data for graphite are taken from Ref.\onlinecite{popvarshneyroy2012}. Temperature evolution of the electrocaloric change in temperature in \BTO/graphite composite under different applied electric fields (b). Inset to panel (b) shows schematically the multilayer geometry of the \BTO\,/graphite composite. }
  \label{Fig3}
\end{figure}

The latter finding also suggests a route to the search or engineering of materials with enhanced electrocaloric response. One example would be a composite of ferroic with a relatively low Curie temperature and a material with high Debye temperature. To predict the potential enhancement of the ECE through such a route we carried out computations of ECE in \BTO/graphite composite. Note, that instead of graphite any other carbon derivative could used.  The choice of  graphite is based on the fact that it has one of the largest Debye temperatures (about 2100~K) \cite{popvarshneyroy2012} and, therefore, is in a quantum regime already at the room temperature. Note, that \BTO/graphene  composites have been successfully synthesized in recent years \cite{LuoACSDielectric2016,WanGraphene201748,WangNanocompositesGraphene}. Here we consider a \BTO\,/graphite composite that has equal volumes of both components. Multilayer geometry for the composite shown schematically in the inset to Fig.\ref{Fig3}(b) allows to avoid electrical conductivity due to graphite. Interestingly, since graphite layers could potentially function as electrodes such geometry could be realized in multilayer capacitors which allow application of  large electric fields. Experimentally in graphene/relaxor polymer nanocomposites electric fields up to 400~kV/cm were applied below the percolation limit \cite{YANG2016461}.
 The volumetric heat capacity of the composite can  be taken as the average value for the two components. Fig.\ref{Fig3}(a) shows the heat capacity of both components of the composite along with the average value for the composite. Since graphite is conducting, the ferroelectric properties of the composite are entirely due to \BTO. To estimate the ECE in the composite we carry out semiclassical simulations of \BTO\, with the heat capacity given by the average value for the composite. Fig.\ref{Fig3}(b) gives the electrocaloric $\Delta T$ for the composite.  We notice a large enhancement of $\Delta T$ as compared to the case of pure \BTO. In particular, under the applied electric field of 650~kV/cm the peak of electrocaloric change in temperature increases from 9.5~K in pure \BTO\,  to 13.9~K in \BTO/graphite composite, while the room temperature value experiences an increase from 2.4 K to 3.8 K. Thus, our simulations confirm that ECE can be enhanced in composites of ferroics with relatively low Curie point and materials with high Debye temperatures. Interestingly, enhancement of the ECE in a relaxor ferroelectric polymer with incorporated  graphene nanofillers was reported  recently \cite{YANG2016461}. It is worth mentioning that the proposed route to the enhanced ECE does not reduce to the heat capacity lowering but rather aims  at increasing the difference between the Debye temperature (or more generally the temperature that marks the transition from quantum to classical behavior) and the Curie temperature.  One possibility would be to lower the Curie temperature with respect to Debye temperature through nanostructuring or epitaxial strain. Another way could be to increase the Debye temperature with respect to Curie temperature by engineering composites of materials with low Curie point and high Debye temperature. 

In summary, we proposed a computational methodology which incorporates temperature dependent heat capacity in adiabatic Monte Carlo simulations. Application of this methodology to direct simulations of ECE in prototypical ferroelectrics allows to quantify the effect that quantum corrections to the heat capacity play in the electrocaloric response of these materials. We find that such corrections always increase the classical estimates for the electrocaloric $\Delta T$ with the effect being most dramatic at low temperatures and in materials where the  Debye temperature exceeds the Curie point. These findings suggest a route to ECE enhancement in composite materials.  

Financial support for this work provided by the National Science
Foundation Grant No. DMR-1250492 and MRI CHE-1531590. The authors
would like to acknowledge the use of the services provided by Research
Computing at the University of South Florida.


\begin{thebibliography}{39}%
\makeatletter
\providecommand \@ifxundefined [1]{%
 \@ifx{#1\undefined}
}%
\providecommand \@ifnum [1]{%
 \ifnum #1\expandafter \@firstoftwo
 \else \expandafter \@secondoftwo
 \fi
}%
\providecommand \@ifx [1]{%
 \ifx #1\expandafter \@firstoftwo
 \else \expandafter \@secondoftwo
 \fi
}%
\providecommand \natexlab [1]{#1}%
\providecommand \enquote  [1]{``#1''}%
\providecommand \bibnamefont  [1]{#1}%
\providecommand \bibfnamefont [1]{#1}%
\providecommand \citenamefont [1]{#1}%
\providecommand \href@noop [0]{\@secondoftwo}%
\providecommand \href [0]{\begingroup \@sanitize@url \@href}%
\providecommand \@href[1]{\@@startlink{#1}\@@href}%
\providecommand \@@href[1]{\endgroup#1\@@endlink}%
\providecommand \@sanitize@url [0]{\catcode `\\12\catcode `\$12\catcode
  `\&12\catcode `\#12\catcode `\^12\catcode `\_12\catcode `\%12\relax}%
\providecommand \@@startlink[1]{}%
\providecommand \@@endlink[0]{}%
\providecommand \url  [0]{\begingroup\@sanitize@url \@url }%
\providecommand \@url [1]{\endgroup\@href {#1}{\urlprefix }}%
\providecommand \urlprefix  [0]{URL }%
\providecommand \Eprint [0]{\href }%
\providecommand \doibase [0]{http://dx.doi.org/}%
\providecommand \selectlanguage [0]{\@gobble}%
\providecommand \bibinfo  [0]{\@secondoftwo}%
\providecommand \bibfield  [0]{\@secondoftwo}%
\providecommand \translation [1]{[#1]}%
\providecommand \BibitemOpen [0]{}%
\providecommand \bibitemStop [0]{}%
\providecommand \bibitemNoStop [0]{.\EOS\space}%
\providecommand \EOS [0]{\spacefactor3000\relax}%
\providecommand \BibitemShut  [1]{\csname bibitem#1\endcsname}%
\let\auto@bib@innerbib\@empty
\bibitem [{\citenamefont {Scott}(2011)}]{Rev-Scott}%
  \BibitemOpen
  \bibfield  {author} {\bibinfo {author} {\bibfnamefont {J.}~\bibnamefont
  {Scott}},\ }\href {\doibase 10.1146/annurev-matsci-062910-100341} {\bibfield
  {journal} {\bibinfo  {journal} {Annu. Rev. Mater. Res.}\ }\textbf {\bibinfo
  {volume} {41}},\ \bibinfo {pages} {229} (\bibinfo {year} {2011})}\BibitemShut
  {NoStop}%
\bibitem [{\citenamefont {Moya}\ \emph {et~al.}(2014)\citenamefont {Moya},
  \citenamefont {Kar-Narayan},\ and\ \citenamefont {Mathur}}]{Rev-Mayo}%
  \BibitemOpen
  \bibfield  {author} {\bibinfo {author} {\bibfnamefont {X.}~\bibnamefont
  {Moya}}, \bibinfo {author} {\bibfnamefont {S.}~\bibnamefont {Kar-Narayan}}, \
  and\ \bibinfo {author} {\bibfnamefont {N.~D.}\ \bibnamefont {Mathur}},\
  }\href {http://dx.doi.org/10.1038/nmat3951} {\bibfield  {journal} {\bibinfo
  {journal} {Nat. Mater}\ }\textbf {\bibinfo {volume} {13}},\ \bibinfo {pages}
  {439} (\bibinfo {year} {2014})}\BibitemShut {NoStop}%
\bibitem [{\citenamefont {Alpay}\ \emph {et~al.}(2014)\citenamefont {Alpay},
  \citenamefont {Mantese}, \citenamefont {Trolier-McKinstry}, \citenamefont
  {Zhang},\ and\ \citenamefont
  {Whatmore}}]{Rev-alpay_mantese_trolier-mckinstry_zhang_whatmore_2014}%
  \BibitemOpen
  \bibfield  {author} {\bibinfo {author} {\bibfnamefont {S.~P.}\ \bibnamefont
  {Alpay}}, \bibinfo {author} {\bibfnamefont {J.}~\bibnamefont {Mantese}},
  \bibinfo {author} {\bibfnamefont {S.}~\bibnamefont {Trolier-McKinstry}},
  \bibinfo {author} {\bibfnamefont {Q.}~\bibnamefont {Zhang}}, \ and\ \bibinfo
  {author} {\bibfnamefont {R.~W.}\ \bibnamefont {Whatmore}},\ }\href {\doibase
  10.1557/mrs.2014.256} {\bibfield  {journal} {\bibinfo  {journal} {MRS
  Bulletin}\ }\textbf {\bibinfo {volume} {39}},\ \bibinfo {pages} {1099–1111}
  (\bibinfo {year} {2014})}\BibitemShut {NoStop}%
\bibitem [{\citenamefont {Valant}(2012)}]{Rev-ValantECEFutureSolid}%
  \BibitemOpen
  \bibfield  {author} {\bibinfo {author} {\bibfnamefont {M.}~\bibnamefont
  {Valant}},\ }\href {\doibase https://doi.org/10.1016/j.pmatsci.2012.02.001}
  {\bibfield  {journal} {\bibinfo  {journal} {Prog. Mater. Sci.}\ }\textbf
  {\bibinfo {volume} {57}},\ \bibinfo {pages} {980 } (\bibinfo {year}
  {2012})}\BibitemShut {NoStop}%
\bibitem [{\citenamefont {Blumenthal}\ and\ \citenamefont
  {Raatz}(2016)}]{Rev-BlumenthalClassificationECECooling}%
  \BibitemOpen
  \bibfield  {author} {\bibinfo {author} {\bibfnamefont {P.}~\bibnamefont
  {Blumenthal}}\ and\ \bibinfo {author} {\bibfnamefont {A.}~\bibnamefont
  {Raatz}},\ }\href {http://stacks.iop.org/0295-5075/115/i=1/a=17004}
  {\bibfield  {journal} {\bibinfo  {journal} {EPL (Europhysics Letters)}\
  }\textbf {\bibinfo {volume} {115}},\ \bibinfo {pages} {17004} (\bibinfo
  {year} {2016})}\BibitemShut {NoStop}%
\bibitem [{\citenamefont {Mischenko}\ \emph {et~al.}(2006)\citenamefont
  {Mischenko}, \citenamefont {Zhang}, \citenamefont {Scott}, \citenamefont
  {Whatmore},\ and\ \citenamefont {Mathur}}]{Rev-MischenkoGiantECEThinFilm}%
  \BibitemOpen
  \bibfield  {author} {\bibinfo {author} {\bibfnamefont {A.~S.}\ \bibnamefont
  {Mischenko}}, \bibinfo {author} {\bibfnamefont {Q.}~\bibnamefont {Zhang}},
  \bibinfo {author} {\bibfnamefont {J.~F.}\ \bibnamefont {Scott}}, \bibinfo
  {author} {\bibfnamefont {R.~W.}\ \bibnamefont {Whatmore}}, \ and\ \bibinfo
  {author} {\bibfnamefont {N.~D.}\ \bibnamefont {Mathur}},\ }\href {\doibase
  10.1126/science.1123811} {\bibfield  {journal} {\bibinfo  {journal}
  {Science}\ }\textbf {\bibinfo {volume} {311}},\ \bibinfo {pages} {1270}
  (\bibinfo {year} {2006})}\BibitemShut {NoStop}%
\bibitem [{\citenamefont {Chang}\ \emph {et~al.}(2015)\citenamefont {Chang},
  \citenamefont {Mani}, \citenamefont {Lisenkov},\ and\ \citenamefont
  {Ponomareva}}]{Rev-ChangThermallyMaechMagnetoelectricPRL}%
  \BibitemOpen
  \bibfield  {author} {\bibinfo {author} {\bibfnamefont {C.-M.}\ \bibnamefont
  {Chang}}, \bibinfo {author} {\bibfnamefont {B.~K.}\ \bibnamefont {Mani}},
  \bibinfo {author} {\bibfnamefont {S.}~\bibnamefont {Lisenkov}}, \ and\
  \bibinfo {author} {\bibfnamefont {I.}~\bibnamefont {Ponomareva}},\ }\href
  {\doibase 10.1103/PhysRevLett.114.177205} {\bibfield  {journal} {\bibinfo
  {journal} {Phys. Rev. Lett.}\ }\textbf {\bibinfo {volume} {114}},\ \bibinfo
  {pages} {177205} (\bibinfo {year} {2015})}\BibitemShut {NoStop}%
\bibitem [{\citenamefont {Liu}\ \emph {et~al.}(2016)\citenamefont {Liu},
  \citenamefont {Scott},\ and\ \citenamefont
  {Dkhil}}]{Rev-YangLiuDirectIndirect}%
  \BibitemOpen
  \bibfield  {author} {\bibinfo {author} {\bibfnamefont {Y.}~\bibnamefont
  {Liu}}, \bibinfo {author} {\bibfnamefont {J.~F.}\ \bibnamefont {Scott}}, \
  and\ \bibinfo {author} {\bibfnamefont {B.}~\bibnamefont {Dkhil}},\ }\href
  {\doibase 10.1063/1.4958327} {\bibfield  {journal} {\bibinfo  {journal}
  {Appl. Phys. Rev.}\ }\textbf {\bibinfo {volume} {3}},\ \bibinfo {pages}
  {031102} (\bibinfo {year} {2016})}\BibitemShut {NoStop}%
\bibitem [{\citenamefont {Lisenkov}\ and\ \citenamefont
  {Ponomareva}(2009{\natexlab{a}})}]{Rev-LisenkovPRBIntrinsicECE}%
  \BibitemOpen
  \bibfield  {author} {\bibinfo {author} {\bibfnamefont {S.}~\bibnamefont
  {Lisenkov}}\ and\ \bibinfo {author} {\bibfnamefont {I.}~\bibnamefont
  {Ponomareva}},\ }\href {\doibase 10.1103/PhysRevB.80.140102} {\bibfield
  {journal} {\bibinfo  {journal} {Phys. Rev. B}\ }\textbf {\bibinfo {volume}
  {80}},\ \bibinfo {pages} {140102} (\bibinfo {year}
  {2009}{\natexlab{a}})}\BibitemShut {NoStop}%
\bibitem [{\citenamefont {Ponomareva}\ and\ \citenamefont
  {Lisenkov}(2012{\natexlab{a}})}]{Rev-PonomarevaPRLBridging}%
  \BibitemOpen
  \bibfield  {author} {\bibinfo {author} {\bibfnamefont {I.}~\bibnamefont
  {Ponomareva}}\ and\ \bibinfo {author} {\bibfnamefont {S.}~\bibnamefont
  {Lisenkov}},\ }\href {\doibase 10.1103/PhysRevLett.108.167604} {\bibfield
  {journal} {\bibinfo  {journal} {Phys. Rev. Lett.}\ }\textbf {\bibinfo
  {volume} {108}},\ \bibinfo {pages} {167604} (\bibinfo {year}
  {2012}{\natexlab{a}})}\BibitemShut {NoStop}%
\bibitem [{\citenamefont {Nishimatsu}\ \emph {et~al.}(2013)\citenamefont
  {Nishimatsu}, \citenamefont {Barr},\ and\ \citenamefont
  {Beckman}}]{Rev-TakeshiDirectMolecularDynamics}%
  \BibitemOpen
  \bibfield  {author} {\bibinfo {author} {\bibfnamefont {T.}~\bibnamefont
  {Nishimatsu}}, \bibinfo {author} {\bibfnamefont {J.~A.}\ \bibnamefont
  {Barr}}, \ and\ \bibinfo {author} {\bibfnamefont {S.~P.}\ \bibnamefont
  {Beckman}},\ }\href {\doibase 10.7566/JPSJ.82.114605} {\bibfield  {journal}
  {\bibinfo  {journal} {J. Phys. Soc. Jpn.}\ }\textbf {\bibinfo {volume}
  {82}},\ \bibinfo {pages} {114605} (\bibinfo {year} {2013})}\BibitemShut
  {NoStop}%
\bibitem [{\citenamefont {Lisenkov}\ \emph {et~al.}(2016)\citenamefont
  {Lisenkov}, \citenamefont {Herchig}, \citenamefont {Patel}, \citenamefont
  {Vaish}, \citenamefont {Cuozzo},\ and\ \citenamefont
  {Ponomareva}}]{Rev-LisenkovECECarbonNano}%
  \BibitemOpen
  \bibfield  {author} {\bibinfo {author} {\bibfnamefont {S.}~\bibnamefont
  {Lisenkov}}, \bibinfo {author} {\bibfnamefont {R.}~\bibnamefont {Herchig}},
  \bibinfo {author} {\bibfnamefont {S.}~\bibnamefont {Patel}}, \bibinfo
  {author} {\bibfnamefont {R.}~\bibnamefont {Vaish}}, \bibinfo {author}
  {\bibfnamefont {J.}~\bibnamefont {Cuozzo}}, \ and\ \bibinfo {author}
  {\bibfnamefont {I.}~\bibnamefont {Ponomareva}},\ }\href {\doibase
  10.1021/acs.nanolett.6b03155} {\bibfield  {journal} {\bibinfo  {journal}
  {Nano Lett.}\ }\textbf {\bibinfo {volume} {16}},\ \bibinfo {pages} {7008}
  (\bibinfo {year} {2016})}\BibitemShut {NoStop}%
\bibitem [{\citenamefont {Lisenkov}\ and\ \citenamefont
  {Ponomareva}(2012)}]{Rev-LisenkovElastocaloricPRB}%
  \BibitemOpen
  \bibfield  {author} {\bibinfo {author} {\bibfnamefont {S.}~\bibnamefont
  {Lisenkov}}\ and\ \bibinfo {author} {\bibfnamefont {I.}~\bibnamefont
  {Ponomareva}},\ }\href {\doibase 10.1103/PhysRevB.86.104103} {\bibfield
  {journal} {\bibinfo  {journal} {Phys. Rev. B}\ }\textbf {\bibinfo {volume}
  {86}},\ \bibinfo {pages} {104103} (\bibinfo {year} {2012})}\BibitemShut
  {NoStop}%
\bibitem [{\citenamefont {Chauhan}\ \emph
  {et~al.}(2015{\natexlab{a}})\citenamefont {Chauhan}, \citenamefont {Patel},\
  and\ \citenamefont {Vaish}}]{Rev-ChauhanECECeramics}%
  \BibitemOpen
  \bibfield  {author} {\bibinfo {author} {\bibfnamefont {A.}~\bibnamefont
  {Chauhan}}, \bibinfo {author} {\bibfnamefont {S.}~\bibnamefont {Patel}}, \
  and\ \bibinfo {author} {\bibfnamefont {R.}~\bibnamefont {Vaish}},\ }\href
  {\doibase 10.1063/1.4919453} {\bibfield  {journal} {\bibinfo  {journal}
  {Appl. Phys. Lett.}\ }\textbf {\bibinfo {volume} {106}},\ \bibinfo {pages}
  {172901} (\bibinfo {year} {2015}{\natexlab{a}})}\BibitemShut {NoStop}%
\bibitem [{\citenamefont {Lisenkov}\ \emph {et~al.}(2013)\citenamefont
  {Lisenkov}, \citenamefont {Mani}, \citenamefont {Chang}, \citenamefont
  {Almand},\ and\ \citenamefont {Ponomareva}}]{Rev-LisenkovMulticaloricPRB}%
  \BibitemOpen
  \bibfield  {author} {\bibinfo {author} {\bibfnamefont {S.}~\bibnamefont
  {Lisenkov}}, \bibinfo {author} {\bibfnamefont {B.~K.}\ \bibnamefont {Mani}},
  \bibinfo {author} {\bibfnamefont {C.-M.}\ \bibnamefont {Chang}}, \bibinfo
  {author} {\bibfnamefont {J.}~\bibnamefont {Almand}}, \ and\ \bibinfo {author}
  {\bibfnamefont {I.}~\bibnamefont {Ponomareva}},\ }\href {\doibase
  10.1103/PhysRevB.87.224101} {\bibfield  {journal} {\bibinfo  {journal} {Phys.
  Rev. B}\ }\textbf {\bibinfo {volume} {87}},\ \bibinfo {pages} {224101}
  (\bibinfo {year} {2013})}\BibitemShut {NoStop}%
\bibitem [{\citenamefont {Chauhan}\ \emph
  {et~al.}(2015{\natexlab{b}})\citenamefont {Chauhan}, \citenamefont {Patel},\
  and\ \citenamefont {Vaish}}]{Rev-ChauhanMulticaloric}%
  \BibitemOpen
  \bibfield  {author} {\bibinfo {author} {\bibfnamefont {A.}~\bibnamefont
  {Chauhan}}, \bibinfo {author} {\bibfnamefont {S.}~\bibnamefont {Patel}}, \
  and\ \bibinfo {author} {\bibfnamefont {R.}~\bibnamefont {Vaish}},\ }\href
  {\doibase https://doi.org/10.1016/j.actamat.2015.01.070} {\bibfield
  {journal} {\bibinfo  {journal} {Acta Mater.}\ }\textbf {\bibinfo {volume}
  {89}},\ \bibinfo {pages} {384 } (\bibinfo {year}
  {2015}{\natexlab{b}})}\BibitemShut {NoStop}%
\bibitem [{\citenamefont {Lisenkov}\ and\ \citenamefont
  {Ponomareva}(2009{\natexlab{b}})}]{Goal-LisenkovPRBIntrinsicECE}%
  \BibitemOpen
  \bibfield  {author} {\bibinfo {author} {\bibfnamefont {S.}~\bibnamefont
  {Lisenkov}}\ and\ \bibinfo {author} {\bibfnamefont {I.}~\bibnamefont
  {Ponomareva}},\ }\href {\doibase 10.1103/PhysRevB.80.140102} {\bibfield
  {journal} {\bibinfo  {journal} {Phys. Rev. B}\ }\textbf {\bibinfo {volume}
  {80}},\ \bibinfo {pages} {140102} (\bibinfo {year}
  {2009}{\natexlab{b}})}\BibitemShut {NoStop}%
\bibitem [{\citenamefont {Ponomareva}\ and\ \citenamefont
  {Lisenkov}(2012{\natexlab{b}})}]{Goal-PonomarevaPRLBridging}%
  \BibitemOpen
  \bibfield  {author} {\bibinfo {author} {\bibfnamefont {I.}~\bibnamefont
  {Ponomareva}}\ and\ \bibinfo {author} {\bibfnamefont {S.}~\bibnamefont
  {Lisenkov}},\ }\href {\doibase 10.1103/PhysRevLett.108.167604} {\bibfield
  {journal} {\bibinfo  {journal} {Phys. Rev. Lett.}\ }\textbf {\bibinfo
  {volume} {108}},\ \bibinfo {pages} {167604} (\bibinfo {year}
  {2012}{\natexlab{b}})}\BibitemShut {NoStop}%
\bibitem [{\citenamefont {Kallaev}\ \emph {et~al.}(2013)\citenamefont
  {Kallaev}, \citenamefont {Omarov}, \citenamefont {Bakmaev},\ and\
  \citenamefont {Abdulvakhidov}}]{Kallaev2013}%
  \BibitemOpen
  \bibfield  {author} {\bibinfo {author} {\bibfnamefont {S.~N.}\ \bibnamefont
  {Kallaev}}, \bibinfo {author} {\bibfnamefont {Z.~M.}\ \bibnamefont {Omarov}},
  \bibinfo {author} {\bibfnamefont {A.~G.}\ \bibnamefont {Bakmaev}}, \ and\
  \bibinfo {author} {\bibfnamefont {K.}~\bibnamefont {Abdulvakhidov}},\ }\href
  {\doibase 10.1134/S1063783413050144} {\bibfield  {journal} {\bibinfo
  {journal} {Phys. Solid State}\ }\textbf {\bibinfo {volume} {55}},\ \bibinfo
  {pages} {1095} (\bibinfo {year} {2013})}\BibitemShut {NoStop}%
\bibitem [{\citenamefont {Rossetti}\ and\ \citenamefont
  {Maffei}(2005)}]{PTOExperimentalRossetti}%
  \BibitemOpen
  \bibfield  {author} {\bibinfo {author} {\bibfnamefont {G.~A.}\ \bibnamefont
  {Rossetti}}\ and\ \bibinfo {author} {\bibfnamefont {N.}~\bibnamefont
  {Maffei}},\ }\href {http://stacks.iop.org/0953-8984/17/i=25/a=021} {\bibfield
   {journal} {\bibinfo  {journal} {J. Phys.: Condens. Matter}\ }\textbf
  {\bibinfo {volume} {17}},\ \bibinfo {pages} {3953} (\bibinfo {year}
  {2005})}\BibitemShut {NoStop}%
\bibitem [{\citenamefont {Todd}\ and\ \citenamefont
  {Lorenson}(1952)}]{doi:10.1021/ja01128a054}%
  \BibitemOpen
  \bibfield  {author} {\bibinfo {author} {\bibfnamefont {S.~S.}\ \bibnamefont
  {Todd}}\ and\ \bibinfo {author} {\bibfnamefont {R.~E.}\ \bibnamefont
  {Lorenson}},\ }\href {\doibase 10.1021/ja01128a054} {\bibfield  {journal}
  {\bibinfo  {journal} {J. Am. Chem. Soc.}\ }\textbf {\bibinfo {volume} {74}},\
  \bibinfo {pages} {2043} (\bibinfo {year} {1952})}\BibitemShut {NoStop}%
\bibitem [{\citenamefont {Wattanasarn}\ and\ \citenamefont
  {Seetawan}(2014)}]{DebyeTempPTO}%
  \BibitemOpen
  \bibfield  {author} {\bibinfo {author} {\bibfnamefont {H.}~\bibnamefont
  {Wattanasarn}}\ and\ \bibinfo {author} {\bibfnamefont {T.}~\bibnamefont
  {Seetawan}},\ }\href {\doibase 10.1080/10584587.2014.905154} {\bibfield
  {journal} {\bibinfo  {journal} {Integr. Ferroelectr.}\ }\textbf {\bibinfo
  {volume} {155}},\ \bibinfo {pages} {59} (\bibinfo {year} {2014})}\BibitemShut
  {NoStop}%
\bibitem [{\citenamefont {Sanna}\ \emph {et~al.}(2011)\citenamefont {Sanna},
  \citenamefont {Thierfelder}, \citenamefont {Wippermann}, \citenamefont
  {Sinha},\ and\ \citenamefont {Schmidt}}]{DebyeTempBTO}%
  \BibitemOpen
  \bibfield  {author} {\bibinfo {author} {\bibfnamefont {S.}~\bibnamefont
  {Sanna}}, \bibinfo {author} {\bibfnamefont {C.}~\bibnamefont {Thierfelder}},
  \bibinfo {author} {\bibfnamefont {S.}~\bibnamefont {Wippermann}}, \bibinfo
  {author} {\bibfnamefont {T.~P.}\ \bibnamefont {Sinha}}, \ and\ \bibinfo
  {author} {\bibfnamefont {W.~G.}\ \bibnamefont {Schmidt}},\ }\href {\doibase
  10.1103/PhysRevB.83.054112} {\bibfield  {journal} {\bibinfo  {journal} {Phys.
  Rev. B}\ }\textbf {\bibinfo {volume} {83}},\ \bibinfo {pages} {054112}
  (\bibinfo {year} {2011})}\BibitemShut {NoStop}%
\bibitem [{Note1()}]{Note1}%
  \BibitemOpen
  \bibinfo {note} {For example, for the case of $<n_{dem}> = x a + (1-x) b$,
  where $a$ and $b$ are the two integers that bracket $<n_{dem}>$ the weighted
  average property $<A>$ is computed as $<A>= x A_a+(1-x)A_b$, where $A_a$ and
  $A_b$ are the property $A$ obtained from simulations with $a$ and $b$ numbers
  of demons, respectively.}\BibitemShut {Stop}%
\bibitem [{Note2()}]{Note2}%
  \BibitemOpen
  \bibinfo {note} {Some data were cross checked by using a simulation supercell
  of 16x16x16 unit cells.}\BibitemShut {Stop}%
\bibitem [{\citenamefont {Mani}\ \emph {et~al.}(2013)\citenamefont {Mani},
  \citenamefont {Chang},\ and\ \citenamefont {Ponomareva}}]{ManiAtomisticPRB}%
  \BibitemOpen
  \bibfield  {author} {\bibinfo {author} {\bibfnamefont {B.~K.}\ \bibnamefont
  {Mani}}, \bibinfo {author} {\bibfnamefont {C.-M.}\ \bibnamefont {Chang}}, \
  and\ \bibinfo {author} {\bibfnamefont {I.}~\bibnamefont {Ponomareva}},\
  }\href {\doibase 10.1103/PhysRevB.88.064306} {\bibfield  {journal} {\bibinfo
  {journal} {Phys. Rev. B}\ }\textbf {\bibinfo {volume} {88}},\ \bibinfo
  {pages} {064306} (\bibinfo {year} {2013})}\BibitemShut {NoStop}%
\bibitem [{\citenamefont {Herchig}\ \emph {et~al.}(2015)\citenamefont
  {Herchig}, \citenamefont {Chang}, \citenamefont {Mani},\ and\ \citenamefont
  {Ponomareva}}]{herchignw}%
  \BibitemOpen
  \bibfield  {author} {\bibinfo {author} {\bibfnamefont {R.}~\bibnamefont
  {Herchig}}, \bibinfo {author} {\bibfnamefont {C.-M.}\ \bibnamefont {Chang}},
  \bibinfo {author} {\bibfnamefont {B.~K.}\ \bibnamefont {Mani}}, \ and\
  \bibinfo {author} {\bibfnamefont {I.}~\bibnamefont {Ponomareva}},\
  }\href@noop {} {\bibfield  {journal} {\bibinfo  {journal} {Sci. Rep.}\
  }\textbf {\bibinfo {volume} {5}},\ \bibinfo {pages} {17294} (\bibinfo {year}
  {2015})}\BibitemShut {NoStop}%
\bibitem [{\citenamefont {Zhong}\ \emph {et~al.}(1995)\citenamefont {Zhong},
  \citenamefont {Vanderbilt},\ and\ \citenamefont
  {Rabe}}]{ZhongPRBHamiltonian}%
  \BibitemOpen
  \bibfield  {author} {\bibinfo {author} {\bibfnamefont {W.}~\bibnamefont
  {Zhong}}, \bibinfo {author} {\bibfnamefont {D.}~\bibnamefont {Vanderbilt}}, \
  and\ \bibinfo {author} {\bibfnamefont {K.~M.}\ \bibnamefont {Rabe}},\ }\href
  {\doibase 10.1103/PhysRevB.52.6301} {\bibfield  {journal} {\bibinfo
  {journal} {Phys. Rev. B}\ }\textbf {\bibinfo {volume} {52}},\ \bibinfo
  {pages} {6301} (\bibinfo {year} {1995})}\BibitemShut {NoStop}%
\bibitem [{\citenamefont {Marathe}\ \emph {et~al.}(2017)\citenamefont
  {Marathe}, \citenamefont {Renggli}, \citenamefont {Sanlialp}, \citenamefont
  {Karabasov}, \citenamefont {Shvartsman}, \citenamefont {Lupascu},
  \citenamefont {Gr\"unebohm},\ and\ \citenamefont
  {Ederer}}]{PhysRevB.96.014102}%
  \BibitemOpen
  \bibfield  {author} {\bibinfo {author} {\bibfnamefont {M.}~\bibnamefont
  {Marathe}}, \bibinfo {author} {\bibfnamefont {D.}~\bibnamefont {Renggli}},
  \bibinfo {author} {\bibfnamefont {M.}~\bibnamefont {Sanlialp}}, \bibinfo
  {author} {\bibfnamefont {M.~O.}\ \bibnamefont {Karabasov}}, \bibinfo {author}
  {\bibfnamefont {V.~V.}\ \bibnamefont {Shvartsman}}, \bibinfo {author}
  {\bibfnamefont {D.~C.}\ \bibnamefont {Lupascu}}, \bibinfo {author}
  {\bibfnamefont {A.}~\bibnamefont {Gr\"unebohm}}, \ and\ \bibinfo {author}
  {\bibfnamefont {C.}~\bibnamefont {Ederer}},\ }\href {\doibase
  10.1103/PhysRevB.96.014102} {\bibfield  {journal} {\bibinfo  {journal} {Phys.
  Rev. B}\ }\textbf {\bibinfo {volume} {96}},\ \bibinfo {pages} {014102}
  (\bibinfo {year} {2017})}\BibitemShut {NoStop}%
\bibitem [{\citenamefont {Tong}\ \emph {et~al.}(2014)\citenamefont {Tong},
  \citenamefont {Karthik}, \citenamefont {Mangalam}, \citenamefont {Martin},\
  and\ \citenamefont {Cahill}}]{PhysRevB.90.094116}%
  \BibitemOpen
  \bibfield  {author} {\bibinfo {author} {\bibfnamefont {T.}~\bibnamefont
  {Tong}}, \bibinfo {author} {\bibfnamefont {J.}~\bibnamefont {Karthik}},
  \bibinfo {author} {\bibfnamefont {R.~V.~K.}\ \bibnamefont {Mangalam}},
  \bibinfo {author} {\bibfnamefont {L.~W.}\ \bibnamefont {Martin}}, \ and\
  \bibinfo {author} {\bibfnamefont {D.~G.}\ \bibnamefont {Cahill}},\ }\href
  {\doibase 10.1103/PhysRevB.90.094116} {\bibfield  {journal} {\bibinfo
  {journal} {Phys. Rev. B}\ }\textbf {\bibinfo {volume} {90}},\ \bibinfo
  {pages} {094116} (\bibinfo {year} {2014})}\BibitemShut {NoStop}%
\bibitem [{\citenamefont {Per\"antie}\ \emph {et~al.}(2010)\citenamefont
  {Per\"antie}, \citenamefont {Hagberg}, \citenamefont {Uusim\"aki},\ and\
  \citenamefont {Jantunen}}]{PhysRevB.82.134119}%
  \BibitemOpen
  \bibfield  {author} {\bibinfo {author} {\bibfnamefont {J.}~\bibnamefont
  {Per\"antie}}, \bibinfo {author} {\bibfnamefont {J.}~\bibnamefont {Hagberg}},
  \bibinfo {author} {\bibfnamefont {A.}~\bibnamefont {Uusim\"aki}}, \ and\
  \bibinfo {author} {\bibfnamefont {H.}~\bibnamefont {Jantunen}},\ }\href
  {\doibase 10.1103/PhysRevB.82.134119} {\bibfield  {journal} {\bibinfo
  {journal} {Phys. Rev. B}\ }\textbf {\bibinfo {volume} {82}},\ \bibinfo
  {pages} {134119} (\bibinfo {year} {2010})}\BibitemShut {NoStop}%
\bibitem [{\citenamefont {Ledbetter}\ \emph {et~al.}(1987)\citenamefont
  {Ledbetter}, \citenamefont {Austin}, \citenamefont {Kim},\ and\ \citenamefont
  {Lei}}]{Ledbetter_Austin_Kim_Lei_1987}%
  \BibitemOpen
  \bibfield  {author} {\bibinfo {author} {\bibfnamefont {H.}~\bibnamefont
  {Ledbetter}}, \bibinfo {author} {\bibfnamefont {M.}~\bibnamefont {Austin}},
  \bibinfo {author} {\bibfnamefont {S.}~\bibnamefont {Kim}}, \ and\ \bibinfo
  {author} {\bibfnamefont {M.}~\bibnamefont {Lei}},\ }\href@noop {} {\bibfield
  {journal} {\bibinfo  {journal} {J. Mat. Res.}\ }\textbf {\bibinfo {volume}
  {2:6}},\ \bibinfo {pages} {786} (\bibinfo {year} {1987})}\BibitemShut
  {NoStop}%
\bibitem [{\citenamefont {Pop}\ \emph {et~al.}(2012)\citenamefont {Pop},
  \citenamefont {Varshney},\ and\ \citenamefont {Roy}}]{popvarshneyroy2012}%
  \BibitemOpen
  \bibfield  {author} {\bibinfo {author} {\bibfnamefont {E.}~\bibnamefont
  {Pop}}, \bibinfo {author} {\bibfnamefont {V.}~\bibnamefont {Varshney}}, \
  and\ \bibinfo {author} {\bibfnamefont {A.~K.}\ \bibnamefont {Roy}},\ }\href
  {\doibase 10.1557/mrs.2012.203} {\bibfield  {journal} {\bibinfo  {journal}
  {MRS Bull.}\ }\textbf {\bibinfo {volume} {37}},\ \bibinfo {pages} {1273}
  (\bibinfo {year} {2012})}\BibitemShut {NoStop}%
\bibitem [{\citenamefont {Luo}\ \emph {et~al.}(2016)\citenamefont {Luo},
  \citenamefont {Wang}, \citenamefont {Tian}, \citenamefont {Gong},
  \citenamefont {Zhao}, \citenamefont {Shen}, \citenamefont {Xu}, \citenamefont
  {Xiao},\ and\ \citenamefont {Li}}]{LuoACSDielectric2016}%
  \BibitemOpen
  \bibfield  {author} {\bibinfo {author} {\bibfnamefont {B.}~\bibnamefont
  {Luo}}, \bibinfo {author} {\bibfnamefont {X.}~\bibnamefont {Wang}}, \bibinfo
  {author} {\bibfnamefont {E.}~\bibnamefont {Tian}}, \bibinfo {author}
  {\bibfnamefont {H.}~\bibnamefont {Gong}}, \bibinfo {author} {\bibfnamefont
  {Q.}~\bibnamefont {Zhao}}, \bibinfo {author} {\bibfnamefont {Z.}~\bibnamefont
  {Shen}}, \bibinfo {author} {\bibfnamefont {Y.}~\bibnamefont {Xu}}, \bibinfo
  {author} {\bibfnamefont {X.}~\bibnamefont {Xiao}}, \ and\ \bibinfo {author}
  {\bibfnamefont {L.}~\bibnamefont {Li}},\ }\href {\doibase
  10.1021/acsami.5b11231} {\bibfield  {journal} {\bibinfo  {journal} {ACS Appl.
  Mater. Inter.}\ }\textbf {\bibinfo {volume} {8}},\ \bibinfo {pages} {3340}
  (\bibinfo {year} {2016})}\BibitemShut {NoStop}%
\bibitem [{\citenamefont {Wan}\ \emph {et~al.}(2017)\citenamefont {Wan},
  \citenamefont {Zhu}, \citenamefont {Yu}, \citenamefont {Yang}, \citenamefont
  {Sun}, \citenamefont {Wong},\ and\ \citenamefont {Liao}}]{WanGraphene201748}%
  \BibitemOpen
  \bibfield  {author} {\bibinfo {author} {\bibfnamefont {Y.-J.}\ \bibnamefont
  {Wan}}, \bibinfo {author} {\bibfnamefont {P.-L.}\ \bibnamefont {Zhu}},
  \bibinfo {author} {\bibfnamefont {S.-H.}\ \bibnamefont {Yu}}, \bibinfo
  {author} {\bibfnamefont {W.-H.}\ \bibnamefont {Yang}}, \bibinfo {author}
  {\bibfnamefont {R.}~\bibnamefont {Sun}}, \bibinfo {author} {\bibfnamefont
  {C.-P.}\ \bibnamefont {Wong}}, \ and\ \bibinfo {author} {\bibfnamefont
  {W.-H.}\ \bibnamefont {Liao}},\ }\href {\doibase
  https://doi.org/10.1016/j.compscitech.2017.01.010} {\bibfield  {journal}
  {\bibinfo  {journal} {Compos. Sci. Technol.}\ }\textbf {\bibinfo {volume}
  {141}},\ \bibinfo {pages} {48 } (\bibinfo {year} {2017})}\BibitemShut
  {NoStop}%
\bibitem [{\citenamefont {Wang}\ \emph {et~al.}(2015)\citenamefont {Wang},
  \citenamefont {Zhu}, \citenamefont {Wang}, \citenamefont {Fan},\ and\
  \citenamefont {Xu}}]{WangNanocompositesGraphene}%
  \BibitemOpen
  \bibfield  {author} {\bibinfo {author} {\bibfnamefont {R.-X.}\ \bibnamefont
  {Wang}}, \bibinfo {author} {\bibfnamefont {Q.}~\bibnamefont {Zhu}}, \bibinfo
  {author} {\bibfnamefont {W.-S.}\ \bibnamefont {Wang}}, \bibinfo {author}
  {\bibfnamefont {C.-M.}\ \bibnamefont {Fan}}, \ and\ \bibinfo {author}
  {\bibfnamefont {A.-W.}\ \bibnamefont {Xu}},\ }\href {\doibase
  10.1039/C4NJ02272F} {\bibfield  {journal} {\bibinfo  {journal} {New J.
  Chem.}\ }\textbf {\bibinfo {volume} {39}},\ \bibinfo {pages} {4407} (\bibinfo
  {year} {2015})}\BibitemShut {NoStop}%
\bibitem [{\citenamefont {Yang}\ \emph {et~al.}(2016)\citenamefont {Yang},
  \citenamefont {Qian}, \citenamefont {Koo}, \citenamefont {Hou}, \citenamefont
  {Zhang}, \citenamefont {Zhou}, \citenamefont {Lin}, \citenamefont {Qiu},\
  and\ \citenamefont {Zhang}}]{YANG2016461}%
  \BibitemOpen
  \bibfield  {author} {\bibinfo {author} {\bibfnamefont {L.}~\bibnamefont
  {Yang}}, \bibinfo {author} {\bibfnamefont {X.}~\bibnamefont {Qian}}, \bibinfo
  {author} {\bibfnamefont {C.}~\bibnamefont {Koo}}, \bibinfo {author}
  {\bibfnamefont {Y.}~\bibnamefont {Hou}}, \bibinfo {author} {\bibfnamefont
  {T.}~\bibnamefont {Zhang}}, \bibinfo {author} {\bibfnamefont
  {Y.}~\bibnamefont {Zhou}}, \bibinfo {author} {\bibfnamefont {M.}~\bibnamefont
  {Lin}}, \bibinfo {author} {\bibfnamefont {J.-H.}\ \bibnamefont {Qiu}}, \ and\
  \bibinfo {author} {\bibfnamefont {Q.}~\bibnamefont {Zhang}},\ }\href
  {\doibase http://dx.doi.org/10.1016/j.nanoen.2016.02.026} {\bibfield
  {journal} {\bibinfo  {journal} {Nano Energy}\ }\textbf {\bibinfo {volume}
  {22}},\ \bibinfo {pages} {461} (\bibinfo {year} {2016})}\BibitemShut
  {NoStop}%
\bibitem [{\citenamefont {Bai}\ \emph {et~al.}(2010)\citenamefont {Bai},
  \citenamefont {Zheng},\ and\ \citenamefont {Shi}}]{ExperimentalBTO}%
  \BibitemOpen
  \bibfield  {author} {\bibinfo {author} {\bibfnamefont {Y.}~\bibnamefont
  {Bai}}, \bibinfo {author} {\bibfnamefont {G.}~\bibnamefont {Zheng}}, \ and\
  \bibinfo {author} {\bibfnamefont {S.}~\bibnamefont {Shi}},\ }\href {\doibase
  10.1063/1.3430045} {\bibfield  {journal} {\bibinfo  {journal} {Appl. Phys.
  Lett.}\ }\textbf {\bibinfo {volume} {96}},\ \bibinfo {pages} {192902}
  (\bibinfo {year} {2010})}\BibitemShut {NoStop}%
\bibitem [{\citenamefont {Kar-Narayan}\ and\ \citenamefont
  {Mathur}(2010)}]{0022-3727-43-3-032002}%
  \BibitemOpen
  \bibfield  {author} {\bibinfo {author} {\bibfnamefont {S.}~\bibnamefont
  {Kar-Narayan}}\ and\ \bibinfo {author} {\bibfnamefont {N.~D.}\ \bibnamefont
  {Mathur}},\ }\href {http://stacks.iop.org/0022-3727/43/i=3/a=032002}
  {\bibfield  {journal} {\bibinfo  {journal} {J. Phys. D: Appl. Phys.}\
  }\textbf {\bibinfo {volume} {43}},\ \bibinfo {pages} {032002} (\bibinfo
  {year} {2010})}\BibitemShut {NoStop}%
\end{thebibliography}

%

%

\newpage 

\end{document}